# Voltage Control of Magnetic Monopoles in Artificial Spin Ice

Andres C. Chavez[1], Anthony Barra[1], and Gregory P. Carman[1]
1. Department of Mechanical and Aerospace Engineering, University of California, Los Angeles, CA 90095, United States of America

Current research on artificial spin ice (ASI) systems has revealed unique hysteretic memory effects and mobile quasi-particle monopoles controlled by externally applied magnetic fields. Here, we numerically demonstrate a strain-mediated multiferroic approach to locally control the ASI monopoles. The magnetization of individual lattice elements is controlled by applying voltage pulses to the piezoelectric layer resulting in strain-induced magnetic precession timed for 180° reorientation. The model demonstrates localized voltage control to move the magnetic monopoles across lattice sites, in CoFeB, Ni, and FeGa based ASI's. The switching is achieved at frequencies near ferromagnetic resonance and requires energies below 620 aJ. The results demonstrate that ASI monopoles can be efficiently and locally controlled with a strain-mediated multiferroic approach.

**Introduction**

Spurred by Pauling's prediction of proton disorder in water ice, spin ices have been widely studied as geometrically frustrated magnetic systems [1–4]. An interesting feature of these materials are quasi-particle magnetic monopoles present in $Ho_2Ti_2O_7$ and $Dy_2Ti_2O_7$ [5–8]. Studies have shown that these monopoles can be manipulated to create magnetic currents analogous to modern electronics which utilize electric charge for storing or propagating information [9–11]. Studying these monopole states is challenging because the relevant phenomena occur at cryogenic temperatures and are sensitive to lattice mismatch between substrate and spin ice crystal [2, 9–11].

To overcome some of these technological challenges, researchers have begun to focus on artificial spin ice (ASI) systems [12–16]. ASI's are lithographically patterned magnetic single domain structures arranged in lattice geometries that are leveraged to create synthetic magnetic monopoles. Pioneering experiments in square and Kagome lattice ASI's have established the presence of magnetic monopoles [12, 17–20]. Stochastic monopole nucleation and movement in ASI's has been achieved by application and subsequent reversal of an externally applied saturating magnetic field [17, 19–22]. Monopole dynamics have been probed through extensive cascade type experiments revealing quenched disorder and freedom from thermal fluctuations [23–25]. Further studies revealed hysteretic memory effects in square lattices following training magnetic field cycles [26]. Although repeatable microstates are achievable, the initial configuration and subsequent limit cycles are stochastically determined. This external field control is problematic since manipulation of magnetic current in ASI's requires individual control of the single magnetic domain structures. Thus, a new approach to manipulate single domain structures in ASI systems is needed for future technological application.

One approach that warrants consideration, to overcome present stochastic ASI control, is strain-mediated multiferroic heterostructures. These multiferroic composites combine present-day piezoelectric and magnetoelastic materials to efficiently control nanoscale magnetism [27–29]. Magnetic control is achieved using voltage-induced piezoelectric strain within patterned magnetoelastic elements [28–32]. Nanopatterned structures resembling the magnetic islands in ASI lattices have been studied on arrays of magnetically dipole-coupled elements with Bennett clocking [33, 34]. The Bennett clocking approach utilizes strain-induced magnetoelastic effects, shape anisotropy, and dipole coupling between neighboring elements to propagate logical information across a lattice. Additionally, numerical and experimental studies have demonstrated strain-induced 180° precessional magnetic switching [35–38]. In this paper, numerical modeling demonstrates magnetic monopole control within a Kagome lattice through switching of the central nodal element using magnetoelastic strain.

**Micromagnetic Simulations of Voltage Control in Artificial Spin Ice**

Our design for controlling magnetic monopoles in a Kagome lattice-based ASI consists of elliptical magnetoelastic islands on a piezoelectric substrate. This design uses voltage induced strains in the piezoelectric layer to control local magnetization states. The magnetoelastic structures are numerically studied with a Landau-Lifshitz-Gilbert (LLG) micromagnetics formulation [39, 40]:

$$\frac{\partial \underline{m}}{\partial t} = -\mu_0 \gamma \left(\underline{m} \times \underline{H}_{eff}\right) + \alpha \left(\underline{m} \times \frac{\partial \underline{m}}{\partial t}\right) \qquad (1)$$

where $\mu_0$ is the permeability of free space, $\gamma$ is the gyromagnetic ratio, and $\alpha$ is the Gilbert damping constant. The model neglects thermal fluctuations and assumes small elastic deformations as well as uniform strains within the magnetoelastic elements [31, 41]. In Equation 1, the effective magnetic field ($\underline{H}_{eff}$) for the ASI system is the sum of the exchange field ($\underline{H}_{ex}$), demagnetization field ($\underline{H}_d$), magnetoelastic field ($\underline{H}_{me}$), and magnetocrystalline anisotropy (MCA) field ($\underline{H}_c$). The magnetoelastic field is defined by $\underline{H}_{me} = 3\lambda_s Y(\varepsilon_x - \varepsilon_y)/\mu_0 M_s$ where $\lambda_s$ is the saturation magnetostriction, $Y$ is the Young's modulus, $M_s$ is the saturation magnetization, and ($\varepsilon_x$, $\varepsilon_x$) are the voltage induced strains in the *x* and *y* directions, respectively. The biaxial strain difference is produced by applying an electric field to patterned electrodes on the piezoelectric layer. The $\underline{H}_{me}$ term is represented in equation (1) by a uniform uniaxial anisotropy defined as $K_\varepsilon = \mu_0 \underline{H}_{me} M_s/2$ [41, 42]. Assuming a cubic crystal structure, the effective field due to MCA is given by $\underline{H}_c^i = -2\left[Kc_1\left(m_j^2 + m_k^2\right) + Kc_2\left(m_j^2 m_k^2\right)\right]/\mu_o M_s$, where $m_i$ is the magnetization component in the *i*th direction and the $Kc_i$ are first and second order cubic anisotropy constants, respectively [29]. This LLG formulation is solved within MuMax3 using a Dormand-Prince finite difference method [43, 44].

In this paper, a four ring multiferroic Kagome lattice ASI containing a monopole was modeled as shown in Figure 1. As seen from the figure, the modeled system is a subset of an infinite Kagome lattice and contains four hexagonal unit cells with a monopole centrally located. Per convention, the dumbbell charge model is used to identify monopoles in an ASI by replacing magnetic dipole moments with two charges (±Q). Specifically, lattice sites in a Kagome ASI exhibiting ±2Q are defined as monopoles. Equivalently an ASI monopole in a Kagome lattice is denoted as a vertex site with the magnetization of three elliptical elements directed into (+2Q) or out (-2Q) of the vertex. The present study assumes the influence of elements beyond the four unit-cells are negligible, so the modeled magnetization switching of the strained element $f_o$ is representative of an infinite lattice. This set of neighboring elements exert the greatest influence on $f_o$ and the resulting dipolar effects on the strain-induced magnetic response can only be captured through simulation of multiple islands. Specifically, it is not possible to know that $f_o$ can switch 180° within the lattice from a simulation of a single ellipse because the dipolar interactions are unknown a priori. Along these lines, the details of the magnetization dynamics (e.g., $\theta \sim atan[m_y/m_x]$) depend on the dipole coupling between neighboring elements for each modeled ASI. The simulated lattice is shown schematically with the hexagonal rings labeled I-IV and the center of each elliptical magnetic island labeled with $f_i$, $r_i$, or $q_i$. The ASI initial magnetization state is chosen with the monopole located at the $f_o$-$r_o$-$q_3$ vertex. Thus, movement of the monopole results from 180° switching of $f_o$'s magnetization so that the magnetization of $f_o$, $r_3$, $q_o$ are directed into their common vertex. The ellipse dimensions were chosen based on a parametric sweep in a micromagnetic simulation focusing on geometries likely to give uniform single domain structures. Ellipses with major axes ranging from 120 nm to 60 nm of ~0.8 aspect ratios were tested. Larger geometries were favored since they will be easier to fabricate in future experiments. Based on the results of the parametric sweep, ellipses with major axes of 100 nm, minor axes of 80 nm, and 3 nm thickness were selected. Per convention, the thermal stability factor at room temperature of modern MRAM devices is set to $40K_bT$. For the materials and geometry chosen, the shape anisotropy of the CoFeB, Ni, and FeGa ellipses are $61K_bT$, $47K_bT$, and $82K_bT$, respectively. The parametric sweep also informed our choice of the separation distance between ellipses, $d$, which was picked to be 25nm. All models used in the parametric sweeps and dynamic studies, which will be presented in the results section, used cuboidal finite-difference-time-domain mesh elements with volumes of 1 nm$^3$. This volume was chosen as a result of convergence studies.

To study dynamic switching of the monopole location within the ASI, precessional magnetic switching of the single domain elliptical magnetoelastic structures was used. This precessional switching was achieved by applying a highly localized voltage-induced strain pulse to a single central ellipse, $f_o$, in the ASI unit cell, in a direction parallel to the minor axis of the ellipse. To confirm that such strain localization is feasible, a finite element study was done to calculate the voltage-controlled strain distribution across the ASI lattice and piezoelectric substrate. The finite element results indicate that the switching strains can be localized to a single ellipse, and nearest neighbors experience, at most, 6% of the threshold switching strain. These results validate the assumption used in our micromagnetic simulation that strains can be localized to an ASI's individual elements. To better represent voltage-induced strain in a working device, the localized strain pulse was ramped linearly to its maximum value. Experimental and numerical studies have demonstrated precessional switching with 70 ps rise/release times of piezostrain. Additional theoretical investigations have suggested that this ramp time can be further reduced by optimizing the electrode geometry, electronics, and piezoelectric film thickness [33, 35–38]. Based on this, the voltage-controlled strain pulses used in this study were programmed to reach a maximum value after 10 ps following a linear ramp function, and they were removed following the same function with a negative slope.

Three magnetoelastic materials were chosen to be included in this modeling study; namely CoFeB, Ni, and FeGa. CoFeB was chosen because of its relatively large saturation magnetization and magnetostriction. Ni was chosen because it is the easiest magnetoelastic material to fabricate. FeGa is chosen for its high magnetoelastic properties which reduce the energy requirements to precessionally switch the single domain elements. The CoFeB and Ni systems are assumed to be polycrystalline with grain sizes much smaller than the exchange length, which allows the MCA to be neglected. The FeGa (19% Ga) is assumed to be single crystal, as the large magnetoelastic response of interest is absent in polycrystalline films. The material properties used for simulation are given in Table I [38, 41, 45–48]. The relevant piezoelectric coefficients of PMN-PT for producing the modeled strain values are found in Ref [49]. A combination of micromagnetic parametric studies and piezoelectric finite element models were used to determine suitable strains, pulse widths, and electric fields for achieving precessional switching of the modeled magnetic elements. The biaxial strain values and pulse widths chosen for CoFeB, Ni, and FeGa were 1600 ppm (4.25 ps), 1600 ppm (3.5 ps), and 2400 ppm (2.7 ps) which correspond to 4.8 MV/m, 4.8 MV/m, and 7.2 MV/m electric fields produced by the electrodes in Figure 1, respectively.

TABLE I. Material properties for modeled artificial spin ice systems.

| Material | $M_s$ (A/m) | $A_{ex}$ (J/m) | A | $\lambda_s$ (ppm) | Y (GPa) | $^a K_{c1}$ (J/m³) | $^b K_{c2}$ (J/m³) |
|---|---|---|---|---|---|---|---|
| CoFeB | 1.2x10⁶ | 1.9x10⁻¹¹ | 0.01 | 50 | 160 | 0 | 0 |
| Ni | 4.8x10⁵ | 1.05x10⁻¹¹ | 0.045 | -34 | 200 | 0 | 0 |
| FeGa | 1.3x10⁶ | 1.4x10⁻¹¹ | 0.04 | 100 | 140 | 1.5x10⁴ | -0.7x10⁵ |

$^a$ Anisotropy vector directed along <1,0,0>
$^b$ Anisotropy vector directed along <0,1,0>

**Results and Discussion**

Figures 2(a)-(c) show strain-induced motion of a monopole defect from the $f_o$-$r_o$-$q_3$ vertex to the $f_o$-$r_3$-$q_o$ vertex in a CoFeB ASI. Figures 2(a)-(b) show the initial (t = 0 ns) and final (t = 3.5 ns) stable magnetic states after strain is applied to $f_o$ at t = 1 ns. The final state in 2(b) shows that the monopole defect was moved from the $f_o$-$r_o$-$q_3$ vertex to the $f_o$-$r_3$-$q_o$ vertex, as evidenced by the 180° rotation of $f_o$ from its initial orientation. The figure also shows that $f_o$'s neighboring lattice elements (both near and far) remain magnetized in their initial direction following strain application. Figure 2(c) provides the rotation angle ($\theta$) as a function of time for $f_o$ and its nearest neighbors ($q_o$, $r_3$, $r_o$, and $q_3$) where the magnetization is taken as a volume average over all the magnetic moments of the ellipse. The non-nearest neighbors are excluded because their magnetization angles are perturbed less than ±5°. In particular, Fig 2(c) shows $f_o$'s magnetization begins to rotate clockwise at t = 1.21 ns. The rotation rate slows slightly at t = 1.4 ns, but continues until the magnetization reaches -180° at t = 1.52 ns. After t = 1.53 ns, $f_o$ overshoots -180° and reaches a maximum rotation angle of -222° at t = 1.6 ns before settling along the -x direction of the ellipse. During the same time, $f_o$'s nearest neighbors remain directed along their initial orientations while experiencing oscillations of less than ~22°. The largest orientation variation for these elements is exhibited by $r_3$ whose magnetization varies between 71° and 50° before settling back to its equilibrium orientation at 60°.

When strain is applied to the central element, $f_o$, of the CoFeB ASI, precessional magnetic switching from 0° to -180° occurs near the ferromagnetic resonance (FMR) frequency of the ellipse. A Fast Fourier transform of the response produces a FMR value of 3.12 GHz with a switching time of 0.32 ns. This FMR value matches closely to the 3.16 GHz FMR predicted by Kittel's equation: $\omega = \gamma \sqrt{(H_{eff} + \mu_0(N_x - N_y)M_y)(H_{eff} + \mu_0(N_z - N_y)M_y)}$ where the effective field is the sum of the mechanical and demagnetization fields. The small difference between the calculated and simulated resonances are attributed to the spatial non-uniformity of the dipolar interactions between $f_o$ and its nearest neighbors in the numerical simulation. In comparison to an isolated ellipse, the symmetric placement of $f_o$'s nearest neighbors makes its major axis magnetically easier during the first 90° of rotation due to the combination of neighbor-derived dipolar and internal shape effects. This positioning raises the energy barrier to rotate $f_o$'s magnetization because any rotation is met with a restoring torque (back towards $\theta = 0$) resulting from nearest neighbor dipolar interactions and internal shape effects. Despite this, the magnetoelastic effects in CoFeB under strain are sufficient to overcome this restoring torque, and push $f_o$ towards a strain-induced easy axis at $\theta = -90°$, as shown in Figure 2(c). As $f_o$ moves to this new easy axis via precessional motion, it is underdamped, and overshoots $\theta = -90°$. However, once the magnetization rotates past -90°, the restoring magnetoelastic torque slows the continued rotation, but this effect vanishes because the timed strain pulse is ramped to zero at t = 1.4 ns. At this moment $f_o$, $q_o$, and $r_3$ all align head on resulting in brief slowing of the rotation at -120°. Although this dipolar interaction slows $f_o$'s rotation, the ellipse's shape anisotropy overcomes this, resulting in continued motion and 180° reorientation.

Figures 3(a)-(b) show dynamic reorientation of the Ni and FeGa ASI lattices, respectively. As observed with CoFeB, the magnetization orientation of the non-nearest neighboring elements remains relatively unchanged following strain. The magnetization of $f_o$ initially rotates clockwise at t = 1.16 ns for the Ni ASI and counterclockwise at t = 1.10 ns for the FeGa ASI. The magnetization rotation slows briefly at t = 1.35 ns and 1.29 ns for the Ni and FeGa ASI's, respectively. As with the CoFeB ASI, ballistic switching occurs and the momentary head-on alignment of $f_o$, $q_o$, and $r_3$ briefly slows rotation at t = 1.35 ns and 1.29 ns for the Ni and FeGa ASI's respectively. However, the rotation continues to -180° at t = 1.53 ns in Ni with 18° of overshoot while the FeGa magnetization rotates to 180° at t = 1.46 ns with 9° of overshoot. As observed in the CoFeB lattice, Figures 3(a)-(b) show that the magnetization of the nearest neighbor elements remains directed along their initial orientations even after $f_o$ rotates 180°. For the Ni system, the largest deviations from equilibrium are exhibited by $r_3$ which varies between 66° and 54° during operation. For the FeGa ASI, $q_3$ experiences the largest shift in magnetic orientation and varies between 110° and 134° during the movement of the monopole.

The magnetization rotation of $f_o$ in both the Ni and FeGa ASI's is qualitatively similar to the CoFeB behavior. However, $f_o$'s magnetization rotates 180° in 0.39 ns at 2.56 GHz for the Ni ASI while it takes 0.35 ns at 2.8 GHz for the FeGa. Similar to the CoFeB system, there is good agreement between the model's resonance and the 2.58 GHz predicted using Kittel's equation. However, the analytical prediction for FeGa of 3.9 GHz is 40% larger than the modeled resonance value. This difference is caused by the incoherent rotation present in the FeGa element's magnetization reorientation, as shown in the panel of Figure 3(b). This only occurs in FeGa elements because the demagnetization energy is much larger than either the CoFeB or Ni elements resulting in spatially non-uniform magnetization rotation. In addition to the comparison of ferromagnetic resonance frequencies, there are other subtle differences in the dynamics of the three lattices. For example, the FeGa system rotates counterclockwise which is caused by the non-deterministic switching present in precessional switching. Another difference is the decrease in overshoot angle and settling times of the Ni and FeGa systems when compared to the CoFeB ASI. This is caused by CoFeB's relatively smaller Gilbert damping coefficient compared Ni and FeGa.

An important observation of the three ASIs is that the flipping of the strained ellipse's moment perturbs the magnetization direction of the neighboring elements, but does not cause them to reorient 180°. Consequently, strain-induced motion of the monopoles does not cause avalanche effects, but gives the capability to deterministically place the ASI in specific magnetic configurations. This precise control is useful for ASI's with return point memory because individual limit cycles can be chosen by explicitly defining the initial state. Alternatively, the microwave properties of spin waves in reprogrammable crystals can be tuned using the control scheme. This is achieved by setting specific magnetic configurations of the ASI lattice which alter the spin wave modes of the crystal.

The amount of energy required to reorient the ASI nanodots can be approximated by evaluating the electrical energy delivered to the piezoelectric layer. For a 100 nm PMN-PT layer, 0.48 V, 0.48 V, and 0.72 V represent the voltages required for CoFeB, Ni, and FeGa respectively. The relative dielectric constant of PMN-PT is $\epsilon = 5880$. Using these values, the amount of electrical energy delivered to the 100 nm x 30 nm electrodes in Figure 1 results in 275 aJ, 275 aJ, and 620 aJ to switch CoFeB, Ni and FeGa, respectively. Despite having larger magnetostrictive properties, the FeGa requires the largest energy to rotate the magnetic moment. This occurs because the ASI geometry is optimized for CoFeB rather than FeGa causing larger demagnetization effects in FeGa compared to the other material systems. Also, the FeGa is not isotropic since it contains an additional magnetocrystalline anisotropy (MCA) that is not included in either the CoFeB or Ni models. Additionally, the suboptimal geometry and MCA for FeGa cause incoherent magnetization rotation in contrast to Ni and CoFeB. Thus, these demagnetization and MCA effects represent additional energy barriers to overcome during precessional switching. Lastly, the energy estimates provided in this manuscript neglect substrate clamping effects, which can be substantial with improperly designed systems. However, this issue has been considered by other researchers (e.g., Ref [38]), suggesting values below 275 aJ are feasible if the system is properly designed.

Experimental implementation of the modeled system is possible with modern micro-fabrication techniques and measurement tools. Specifically, the ASI can be defined using electron beam lithography and local strain can be generated by applying voltage to the patterned electrodes of Figure 1. Device production, however is non-trivial and considerable attention must be paid to the layout of electrodes, interconnects, and testing interfaces. Simultaneous control of multiple monopoles is necessary for practical device implementation and, although difficult, it is possible with contemporary CMOS control systems/software.

In this study, a multiferroic artificial spin ice design was proposed for local magnetization control of magnetic monopoles. Strain-mediated monopole movement in a Kagome lattice ASI was demonstrated numerically for CoFeB, Ni, and FeGa systems with motion achieved through precessional magnetization switching near FMR. In contrast to contemporary stochastic control methods, the results demonstrate that ASI monopoles can be efficiently and locally controlled using a strain-mediated multiferroic approach. This demonstration provides a control methodology for future ASI based technologies requiring deterministic manipulation of monopole states.

ACKNOWLEDGEMENT: This project was supported in part by FAME, one of six centers of STARnet, a Semiconductor Research Corporation program sponsored by MARCO and DARPA. Additionally, this work was supported by the NSF Nanosystems Engineering Research Center for Translational Applications of Nanoscale Multiferroic Systems (TANMS) under the Cooperative Agreement Award EEC-1160504.

## References

[1] L. Pauling, "The Structure and Entropy of Ice and of Other Crystals with Some Randomness of Atomic Arrangement," *J. Am. Chem. Soc.*, vol. 57, no. 12, pp. 2680–2684, 1935.

[2] J. E. Greedan, "Frustrated rare earth magnetism: Spin glasses, spin liquids and spin ices in pyrochlore oxides," *J. Alloys Compd.*, vol. 408–412, pp. 444–455, 2006.

[3] B. C. den Hertog and M. J. P. Gingras, "Dipolar Interactions and Origin of Spin Ice in Ising Pyrochlore Magnets," 2000.

[4] S. T. Bramwell and M. J.-P. Gingras, "Spin Ice State in Frustrated Magnetic Pyrochlore Materials," vol. 294, no. November, pp. 1495–1502, 2002.

[5] T. Fennell, P. P. Deen, A. R. Wildes, K. Schmalzl, D. Prabhakaran, A. T. Boothroyd, R. J. Aldus, D. F. McMorrow, and S. T. Bramwell, "Magnetic Coulomb phase in the spin ice $Ho_2Ti_2O_7$," *Science (80-. ).*, vol. 326, no. 5951, pp. 415–417, 2009.

[6] D. J. P. Morris *et al.*, "Dirac Strings and Magnetic Monopoles in the Spin Ice $Dy_2Ti_2O_7$," *Science (80-. ).*, vol. 326, no. 5951, pp. 411–414, 2009.

[7] L. D. C. Jaubert and P. C. W. Holdsworth, "Signature of magnetic monopole and Dirac string dynamics in spin ice," *Nat. Phys.*, vol. 5, no. 4, pp. 258–261, 2009.

[8] C. Castelnovo, R. Moessner, and S. L. Sondhi, "Magnetic monopoles in spin ice," *Nature*, vol. 451, no. 7174, pp. 42–45, 2008.

[9] S. R. Giblin, S. T. Bramwell, P. C. W. Holdsworth, D. Prabhakaran, and I. Terry, "Creation and measurement of long-lived magnetic monopole currents in spin ice," *Nat. Phys.*, vol. 7, no. 3, pp. 252–258, 2011.

[10] S. J. Blundell, "Monopoles, magnetricity, and the stray field from spin ice," *Phys. Rev. Lett.*, vol. 108, no. 14, pp. 1–5, 2012.

[11] S. T. Bramwell, S. R. Giblin, S. Calder, R. Aldus, D. Prabhakaran, and T. Fennell, "Measurement of the charge and current of magnetic monopoles in spin ice," *Nature*, vol. 461, no. 7266, pp. 956–959, 2009.

[12] S. Zhang, "Tuning Geometries and Interactions of Artificial Frustrated Nanomagnets," no. October, 2013.

[13] I. Gilbert, B. R. Ilic, B. Robert Ilic, and B. R. Ilic, "Exploring frustrated magnetism with artificial spin ice," *SPIE 9931, Spintron. IX, 99311P*, vol. 9931, p. 99311P, 2016.

[14] K. K. Kohli, A. L. Balk, J. Li, S. Zhang, I. Gilbert, P. E. Lammert, V. H. Crespi, P. Schiffer, and N. Samarth, "Magneto-optical Kerr Effect Studies of Square Artificial Spin Ice," *Phys. Rev. B*, vol. 84, no. 18, pp. 1–5, 2011.

[15] R. F. Wang *et al.*, "Artificial 'spin ice' in a geometrically frustrated lattice of nanoscale ferromagnetic islands," *Nature*, vol. 439, no. 7074, pp. 303–306, 2006.

[16] C. Nisoli, R. Moessner, and P. Schiffer, "Colloquium: Artificial spin ice: Designing and imaging magnetic frustration," *Rev. Mod. Phys.*, vol. 85, no. 4, pp. 1473–1490, 2013.

[17] S. Ladak, D. E. Read, W. R. Branford, and L. F. Cohen, "Direct observation and control of magnetic monopole defects in an artificial spin-ice material," *New J. Phys.*, vol. 13, no. 5, pp. 359–363, 2011.

[18] K. Zeissler, S. K. Walton, S. Ladak, D. E. Read, T. Tyliszczak, L. F. Cohen, and W. R. Branford, "The non-random walk of chiral magnetic charge carriers in artificial spin ice.," *Sci. Rep.*, vol. 3, p. 1252, 2013.

[19] E. Mengotti, L. J. Heyderman, A. F. Rodríguez, F. Nolting, R. V. Hügli, and H.-B. Braun, "Real-space observation of emergent magnetic monopoles and associated Dirac strings in
7

**Figures**

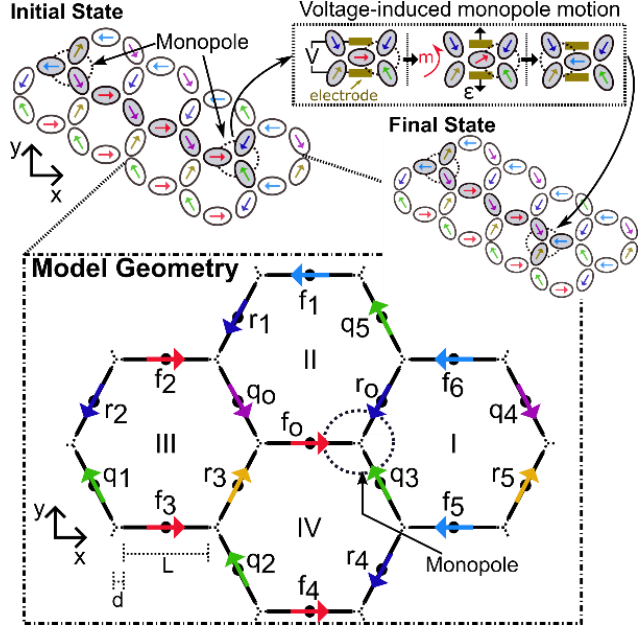

Figure 1. (a) Schematic illustrating monopole defect motion in Kagome ASI caused by voltage-induced strain. Inset shows the initial magnetic orientation of the four rings modeled with a monopole defect located at $f_o$-$r_o$-$q_3$ vertex.

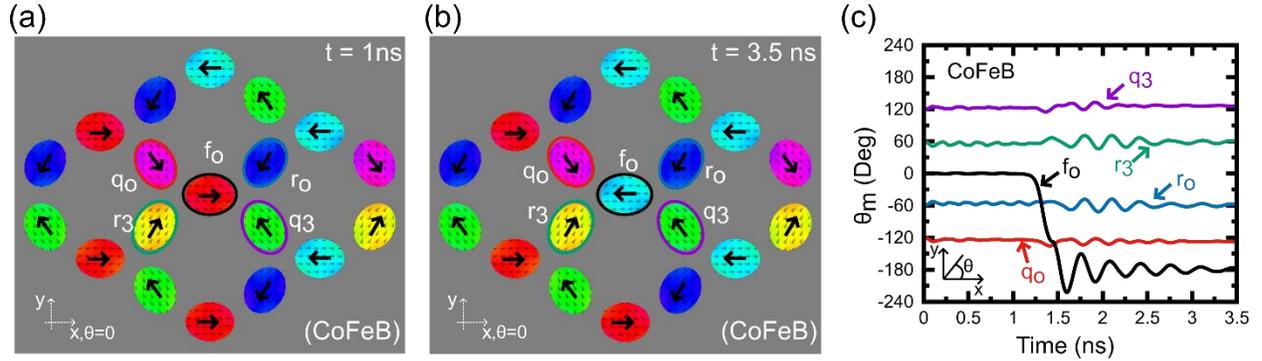

Figure 2. (a) Magnetic state of CoFeB ASI after relaxing for 1 ns. (b) Final state of CoFeB ASI after application of strain pulse. (c) Time variation of x-component of magnetization for central magnetic island and nearest neighbors.

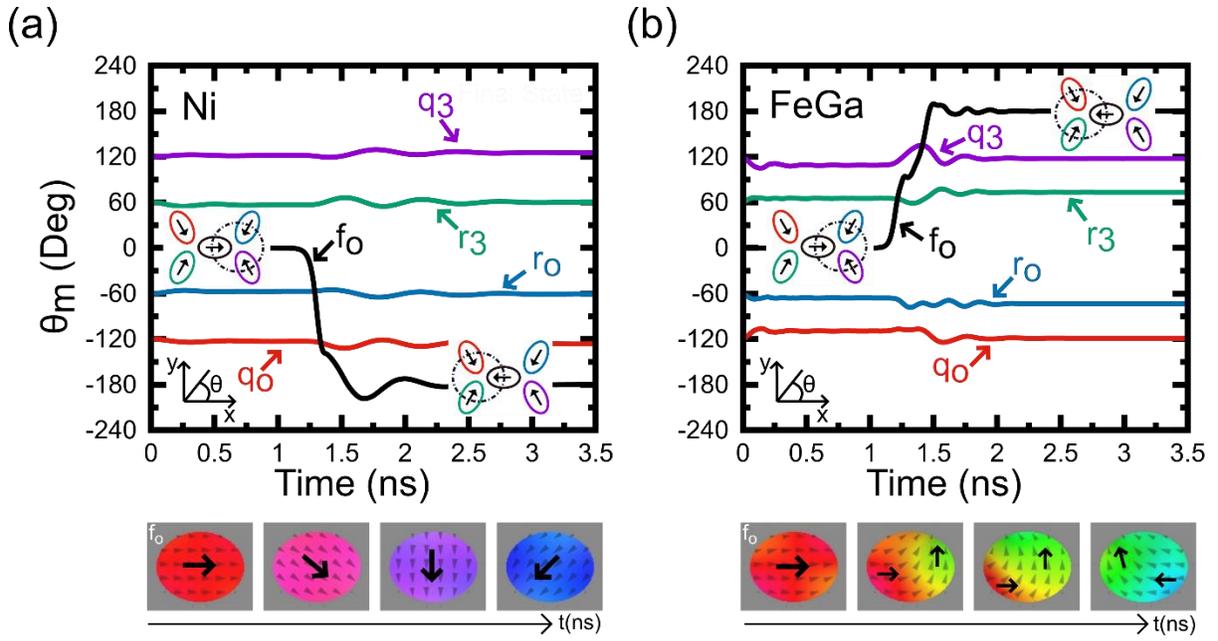

Figure 3. Time variation of x-component of magnetization for central magnetic island and nearest neighbors of (a) Ni ASI system and (b) FeGa ASI system.